%% file: 00-main.tex
\documentclass[sigconf]{acmart}
\input 01-packages
\input 02-macros

\graphicspath{{images/}}

\copyrightyear{2023} 
\acmYear{2023} 
\setcopyright{acmlicensed}\acmConference[SIGGRAPH '23 Conference Proceedings]{Special Interest Group on Computer Graphics and Interactive Techniques Conference Conference Proceedings}{August 6--10, 2023}{Los Angeles, CA, USA}
\acmBooktitle{Special Interest Group on Computer Graphics and Interactive Techniques Conference Conference Proceedings (SIGGRAPH '23 Conference Proceedings), August 6--10, 2023, Los Angeles, CA, USA}
\acmPrice{15.00}
\acmDOI{10.1145/3588432.3591547}
\acmISBN{979-8-4007-0159-7/23/08}

\begin{document}

\author{Ryan Capouellez}
\affiliation{%
  \institution{New York University}
  \country{USA}
}
\email{rjc8237@nyu.edu}

\author{Jiacheng Dai}
\affiliation{%
  \institution{New York University}
  \country{USA}
}
\email{jd4705@nyu.edu}

\author{Aaron Hertzmann}
\affiliation{%
  \institution{Adobe}
  \country{USA}
}
\email{hertzman@dgp.toronto.edu}

\author{Denis Zorin}
\affiliation{%
  \institution{New York University}
  \country{USA}
}
\email{dzorin@cs.nyu.edu}

\title{Algebraic Smooth Occluding Contours}
\begin{abstract}
Computing occluding contours is a key step in 3D non-photorealistic rendering, but producing smooth contours with consistent visibility has been a notoriously-challenging open problem. This paper describes the first general-purpose smooth surface construction for which the occluding contours can be computed in closed form. Given an input mesh and camera viewpoint, we show how to approximate the mesh with a $G^1$ piecewise-quadratic surface, for which the occluding contours are piecewise-rational curves in image-space. We show that this method produces smooth contours with consistent visibility much more efficiently than the state-of-the-art.
\end{abstract}

\ccsdesc[500]{Computing methodologies~Non-photorealistic rendering}
\ccsdesc[500]{Computing methodologies~Visibility}
\ccsdesc[500]{Computing methodologies~Parametric curve and surface models}

\keywords{contours, non-photorealistic rendering, piecewise-quadratic surface, visibility}

\input{figure_tex/teaser.tex}

\maketitle

\input 10-intro

\input 20-related

\input 30-overview
\input 35-extraction

\input 40-construction

\input 50-results
\input 60-concl

\bibliographystyle{ACM-Reference-Format}
\bibliography{contours}

\input 65-figures

%
%


%
%


\end{document}


\title{Algebraic Smooth Occluding Contours: Supplemental material}

\author{Ryan Capouellez, Jiacheng Dai, Aaron Hertzmann, and Denis Zorin}
\date{}

\maketitle
\input{70-suppl.tex}

%% file: 01-packages.tex
\usepackage[ruled,vlined,noend]{algorithm2e}
\usepackage[utf8]{inputenc}
\usepackage{amsmath}
\usepackage{amsfonts}
\usepackage{amsthm}
\usepackage{graphicx}
\usepackage{xcolor}

\usepackage{multirow}

%% file: 02-macros.tex
\iftrue 
\newcommand{\denis}[1]{\textcolor{blue}{\bf [DZ: #1]}}
\newcommand{\DZ}[1]{\textcolor{blue}{\bf [DZ: #1]}}
\newcommand{\aaron}[1]{\textcolor{red}{\bf [AH: #1]}}
\newcommand{\ryan}[1]{\textcolor{magenta}{\bf [RC: #1]}}
\newcommand{\jiacheng}[1]{\textcolor{cyan}{\bf [JD: #1]}}

\else
\newcommand{\denis}[1]{}
\newcommand{\DZ}[1]{}
\newcommand{\aaron}[1]{}
\newcommand{\ryan}[1]{}
\newcommand{\jiacheng}[1]{}

\fi

\SetAlFnt{\small}
\SetAlCapFnt{\small}
\SetAlCapNameFnt{\small}
\SetAlCapHSkip{0pt}

\newcommand{\bR}{\mathbb{R}}
\newcommand{\bp}{\mathbf{p}}

\newcommand{\bn}{\mathbf{n}}
\newcommand{\bb}{\mathbf{b}}
\newcommand{\bc}{\mathbf{c}}
\newcommand{\bt}{\mathbf{t}}

\newcommand{\br}{\mathbf{r}}

\newcommand{\be}{\mathbf{e}}
\newcommand{\bg}{\mathbf{g}}

\newcommand{\bz}{\mathbf{z}}

\newcommand{\bP}{\mathbf{P}}

\newtheorem{prop}{Proposition}

\AtBeginDocument{%
  \providecommand\BibTeX{{%
    \normalfont B\kern-0.5em{\scshape i\kern-0.25em b}\kern-0.8em\TeX}}}

%% file: figure_tex/teaser.tex
\begin{teaserfigure}
\centering
\includegraphics[width=\textwidth]{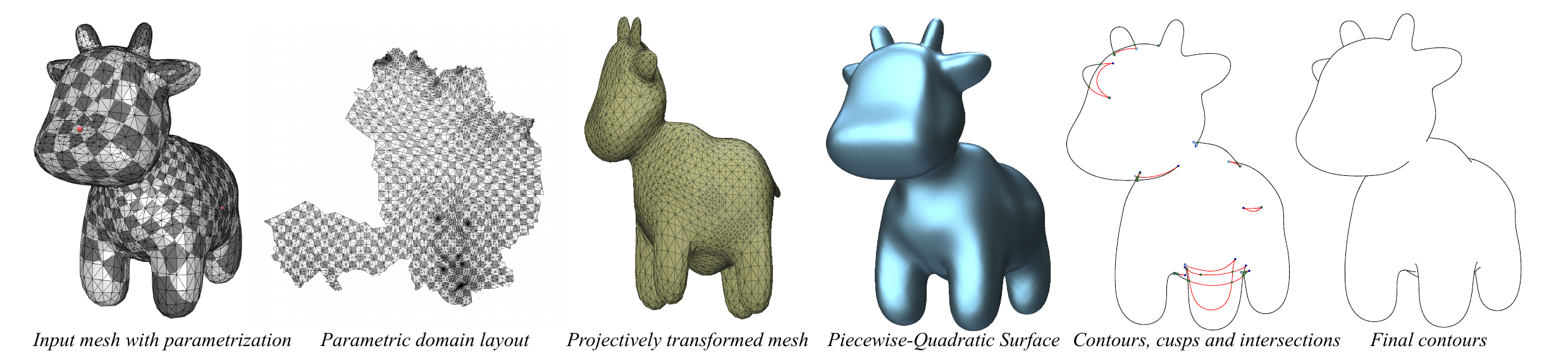}
\caption{Our
method takes a triangle mesh, and renders smooth occluding contours for a given camera viewpoint. 
For each view, the algorithm produces a piecewise-quadratic surface, for which we show the occluding contours can be computed algebraically under orthographic projection. We use a projective transformation of the input mesh to handle perspective projection. By precomputing a conformal parameterization and a mapping from the viewpoint to the quadratic coefficients, contours and visibility can be computed very efficiently, in contrast to previous methods that used expensive heuristics or computations to produce  smooth contours with accurate visibility. (Public domain Spot model by Keenan Crane.)
\label{fig:teaser}
}
\end{teaserfigure}

%% file: 10-intro.tex
\section{Introduction}
\label{sec:intro}

Computing occluding contour lines of 3D objects is a common step in 3D non-photorealistic rendering algorithms, whether for  architectural drawings, cartoon stylization, or pen-and-ink illustration. Conceptually, the problem is deceptively simple: find the points of the surface where the dot product $\bn \cdot \tau$ of the view direction $\tau$ and the normal vector $\bn$ changes sign, and determine which of these points are visible \cite{BenardHertzmann}. 

For triangle meshes, the exact occluding contours can easily be defined: they are a subset of the edges of the original mesh. However, when the mesh represents a smooth object, these contours usually have many spurious singularities, and do not produce the clean contour topology of a smooth surface (Figure~\ref{fig:mesh_contours}). This makes them unsuitable for curve stylization, and noisy during animation, e.g., \cite{Benard:2014}. 
This mismatch between mesh contours and smooth surface contours can be attributed  to the fact that the normal $\bn$ is discontinuous on the mesh, and sign-change sets of discontinuous functions fundamentally differ in structure from the zero sets of smooth functions.

Unfortunately, robustly computing the occluding contours for smooth surfaces is difficult. For common representations, the occluding contours are projective-transformed piecewise higher-order algebraic implicit curves. Existing methods approximate these curves with polylines, but visibility for these polylines is unreliable, as they are not the contours of the smooth surface \cite{Eisemann:2008,Benard:2014}. Recent methods construct a new triangle mesh with the extracted polylines as contours \cite{Benard:2014,ConTesse}, but require very costly heuristic search. These difficulties raise the question: is there a practical smooth representation for which visible occluding contours can be computed exactly?

In this paper, we describe a method for computing occluding contours in \emph{closed form}. For a given input mesh and camera viewpoint, we show how to approximate the mesh with a $C^1$ (excluding a small set of points) surface so that the occluding contours are piecewise-rational curves in image space. This algebraic representation allows for reliable visibility by solving low-order polynomial equations, and for direct, efficient computations without heuristics or search.


Our approach operates as follows. We first apply a projective transformation to the input mesh, reducing the problem to orthographic occluding contours. We use a Powell-Sabin construction to produce a $C^1$ \emph{quadratic patch} surface, which, we observe, is the only algebraic representation with rational contours. To support perspective views, our surface construction is \emph{view-dependent} but can be evaluated very efficiently with a precomputed matrix factorization. All possible piecewise rational quadratic contour lines in the parametric domain are easily enumerated, and correspond to piecewise rational quartic curves in the image domain.   
Visibility for the contours is also resolved precisely by a straightforward computation (up to numerical errors in solving low-order polynomial equations). 

In proposing the first algebraic smooth occluding contour procedure, our main contribution is a careful integration and adaptation of a number of recent and classical techniques in a highly-constrained setting. This includes recent methods for robust surface parameteriztion, a  view-dependent almost-everywhere $C^1$ surface approximation, supporting efficient per-view updates with precomputation, quantitative invisibility  for piecewise quadratic surfaces, and efficient cusp computation.

In comparison to  state-of-the-art methods for accurate contours \cite{Benard:2014,ConTesse}, our method does not involve expensive search and unpredictable refinement heuristics; we find order-of-magnitude faster performance on larger meshes, while avoiding the visibility errors that older methods are prone to.

%% file: 20-related.tex

\section{Related Work}
\label{sec:related}
It has long been known that the occluding contours for a smooth surface cannot be computed analytically, except for simple primitives like spheres and quadrics \cite{Cipolla:2000}. 
Hence, all previous methods employ numerical approximations of the smooth contour; see \cite{BenardHertzmann} for a survey.
Past methods either often produce artifacts, or else require expensive computations to achieve topologically-accurate curves.

Raster methods, based on edge detection of an image buffer are the simplest way to approximate the contour,
e.g., \cite{Saito:1990,Decaudin:1996}; however, these methods do not produce a vector representation of the contours. 

Representing a smooth surface as a triangle mesh yields accurate image-space contours but with erroneous topology (Figure \ref{fig:mesh_contours}).
Heuristics may be used to smooth the topology for stylization \cite{Eisemann:2008,Isenberg:2002,Kirsanov:2003,Northrup:2000}, but these too may be very inaccurate and often produce artifacts. 

\input{figure_tex/mesh_contours.tex}

A third approach is to directly compute a polyline contour approximation, e.g., by root-finding on the smooth surface representation \cite{Weiss:1966:VPI:321328.321330,Elber:1990,Stroila:2008,Winkenbach:1996}.
 However, visibility of these sampled contours are inconsistent with the smooth surface \cite{Benard:2014}. Defining a piecewise-smooth contour function on a triangle mesh \cite{Hertzmann:2000} has the same problem. B\'{e}nard et al.~\shortcite{Benard:2014} generate a new triangle mesh from sampled contours that produces consistent visibility.  However, this method has a high computational cost and is not guaranteed to find a valid mesh.  Planar-maps can produce consistent visibility  \cite{Stroila:2008,Winkenbach:1996}, but may also include incorrect topology from polyline contour approximations.

Liu et al.~\shortcite{ConTesse} recently explained why this problem has been so difficult: sampling smooth contours produces 2D polylines that cannot be the contours of any valid surface. Liu et al.~do guarantee valid polylines, but their method uses expensive numerical sampling operations, involving costly iterative refinements and heuristics to find a consistent mesh.

In contrast to these works, we present a method using a $C^1$ approximation of an arbitrary input mesh for which we can compute \emph{exact} (up to numerical precision) contours algebraically, with precise visibility, without any of the heuristics or expensive refinement procedures of previous methods.  

The geometric modeling literature on constructing $G^1/C^1$ surfaces is vast, but due to constraints of our problem, only a few constructions are relevant in our context; 
\cite{he2005c,powell1977piecewise} which we rely on, and \cite{dahmen1989smooth} being most closely related. We provide more details in Section~\ref{sec:surface}. 

\emph{Linear-normal} (LN) surfaces  
\cite{jue:98b,juttler2000hermite} produce $G^1$ surfaces with normals depending linearly on the parametric coordinates, which, in turn, yields piecewise-linear contour lines in the parametric domain, but they are limited in the type of surfaces that these can represent, e.g., they have singularities at parabolic points.


%% file: figure_tex/mesh_contours.tex
\begin{figure}
\centering
    (a)
\includegraphics[width=1.5in]{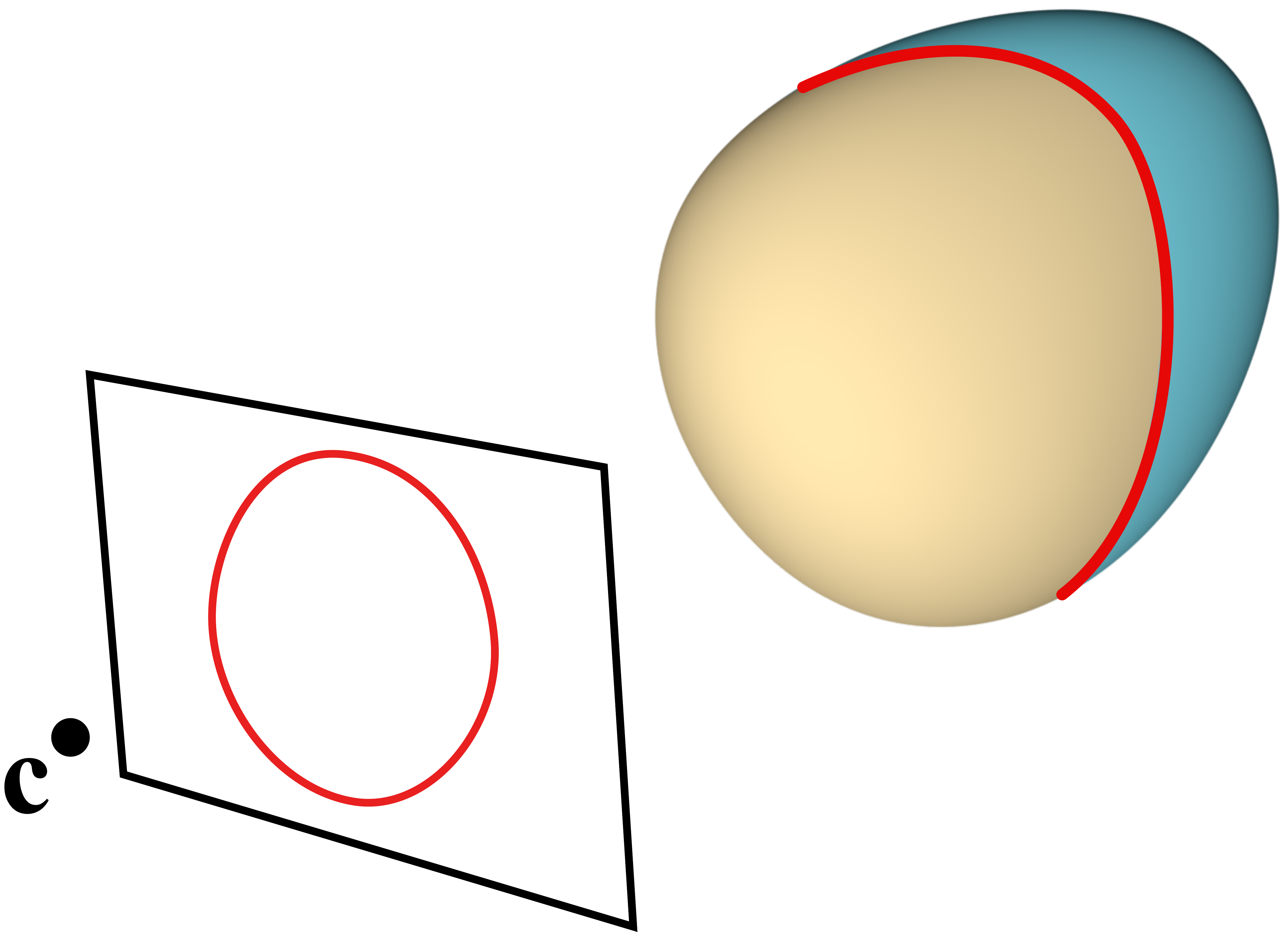}%
~%
(b)
\includegraphics[width=1.5in]{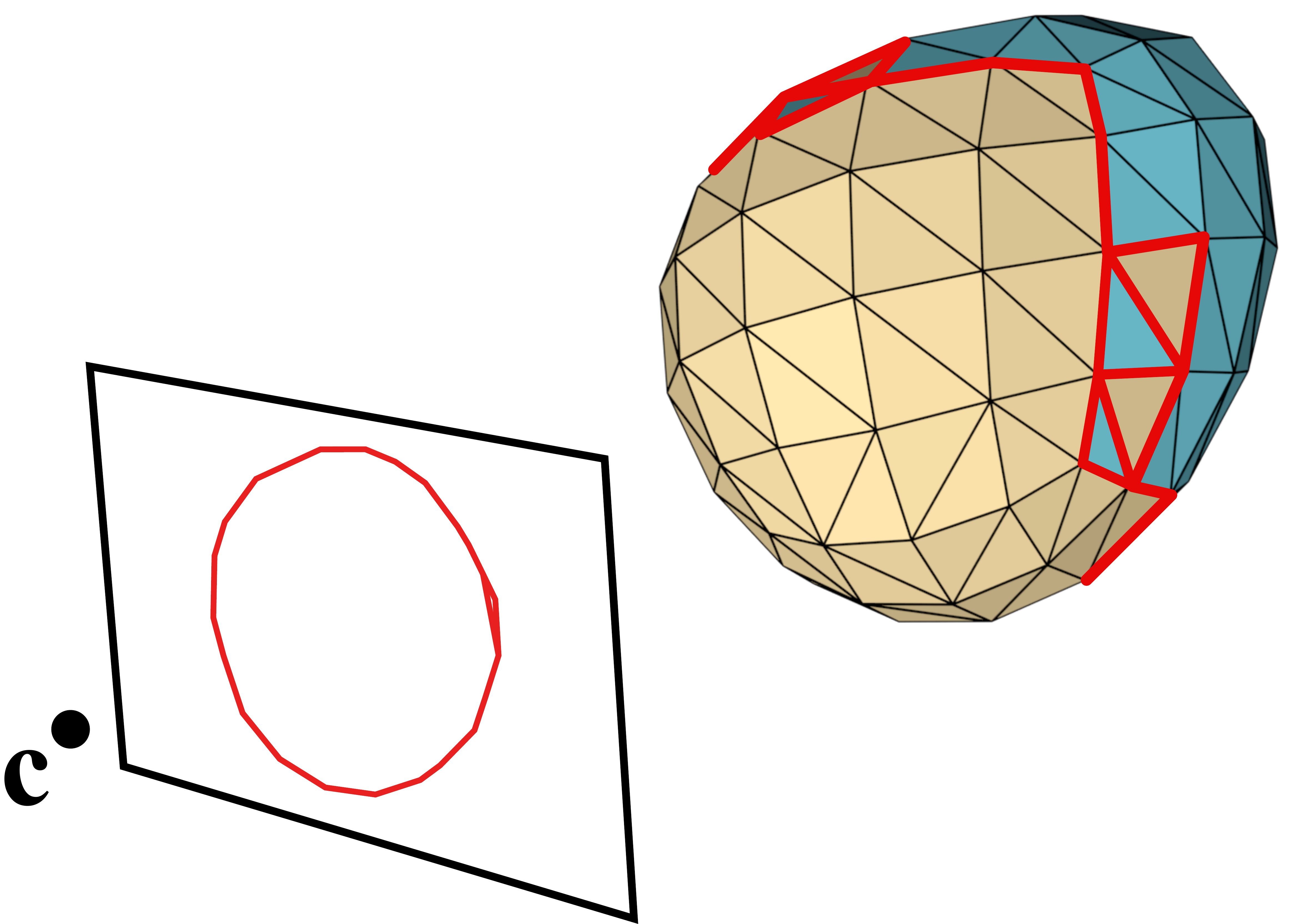} 
\caption{
While contour lines on piecewise-linear surfaces can be computed robustly and efficiently, these consist of a noisy set of mesh edges (b), not suitable for constructing, e.g., vector approximations of contours or stylized rendering;  in comparison, smooth surface contours (a) typically yield a clean spatial curve.  
Figure courtesy of the authors of \cite{ConTesse}. 
\label{fig:mesh_contours}
}
\end{figure}

%% file: 30-overview.tex
\section{Overview}
\label{sec:overview}

Given an input mesh $M$ and camera position $\bc$, we seek a tangent plane-continuous surface representation that approximates the mesh and yields high-quality contours that are continuous and have explicit algebraic form, allowing exact and efficient computation. 
Our key observation is that the problem of finding this representation is extremely constrained---yet it does have a solution. We start with defining the problem more precisely. 

  We use $\bp(u,v): \Omega \subset \bR^2 \rightarrow \bR^3$ to denote 
a  surface patch parameterized over a domain $\Omega$ in the plane. The parameterization is assumed to be at least $C^1$.
The normal $\bn(u,v) = \bp_u \times \bp_v$  is not necessarily unit length.
The camera position is  $\bc$, and the image plane is $P$, with  unit normal $\tau$. 

The  \emph{occluding contour} of the  image of the surface patch $\bp$ in $P$ is the projection to the image plane of  the set of points $\bp(C)$, with the curve $C$ in the parametric domain defined by 
\begin{equation}
    \bn(u,v) \cdot (\bc - \bp(u,v)) = 0\;\mbox{(perspective),}\;  \bn(u,v) \cdot \tau = 0\;\mbox{(orthographic)}
    \label{eq:contour-eq}
\end{equation}
The curve $\bp(C)$ is the \emph{occluding contour generator}, and the apparent contour is the visible projection of that curve. 
For brevity, we refer to each of these 2D and 3D curves as ``contours.''  These elements are illustrated in Figure \ref{fig:patch_contours}.  Existing methods for solving these equations employ numerical approximations, as no closed-form solution is known for these equations for general-purpose smooth surfaces.

\input{figure_tex/patch_contours.tex}

\paragraph{Algebraic Contour Existence}
Under what conditions does Equation \ref{eq:contour-eq} yield curves in closed-form?  
For algebraic functions, the following is known, e.g., \cite{ferrer2008rational}:
\begin{prop}
An irreducible algebraic curve $c(u,v) = 0$  admits a rational parameterization for an arbitrary choice of coefficients if and only if it is linear or quadratic. 
\end{prop}

The linear case corresponds to using a triangle mesh, which has been heavily used in previous methods, with the problems discussed in Section \ref{sec:related}. Juttler et al. \cite{jue:98b,juttler2000hermite} describe a higher-order algebraic surface with constraints that make $C$ reducible into a linear factor and another factor independent of the view direction, but the construction is degenerate at parabolic points. 
Instead, we consider surfaces where $\bp(u,v)$ is \emph{quadratic}, which has not been explored in depth. When $\bp(u,v)$ is quadratic, $\bn(u,v)$ is also quadratic.
Then, for orthographic projection, Equation \ref{eq:contour-eq} is quadratic, and thus the solution curves are rational functions.  It is also clear from Equation \ref{eq:contour-eq} that contour continuity requires that the surface is at least $G^1$.

\emph{ Hence, if we can approximate an input mesh with a $G^1$, piecewise-quadratic surface, then its contours are piecewise rational functions for orthographic projections. }

Next, we summarize the key features of our method, determined by this observation. 

\paragraph{Handling Perspective.}
Many applications require perspective projection, which we handle by applying a projective transformation to the input mesh. That is, an input vertex with camera coordinates $[x,y,z]$ becomes $[x/z, y/z, -1/z]$, with orthographic view vector $\tau=[0,0,1]$, yielding a view equivalent to perspective projection of the original input.  A key insight is that, for algebraic contours, we must apply the transformation to the mesh
\textit{before} constructing a smooth representation, which makes the surface approximation view-dependent, but does not lead to visible artifacts.

\paragraph{Quadratic Surface Construction}
After the projective transformation, we seek to approximate the mesh with a surface composed of quadratic patches. In order for the contours to be continuous across patch boundaries, the surface must be $G^1$  everywhere, except a small number of isolated points. The existing surface choices are quite constrained.
A classical solution to this problem for surfaces parameterized over arbitrary triangulations of the plane is the \emph{Powell-Sabin} interpolant, which was in part motivated by a need for continuous isolines for height fields.
Applying these to arbitrary meshes requires several additional components: mesh parametrization, a method for dealing with singular vertices of the parametrization, a way to generate high-quality surfaces without extraneous oscillations, and fast update to handle view-dependence.  Our solution is based 
on the overall idea of He et  al~\shortcite{he2005c}, but with several 
important differences.  In particular, their method does not produce quadratic patches for one-ring of triangles near singular vertices of the parametrization (cones) and it is interpolating, rather than approximating, the input mesh. 



\paragraph{Visibility} 
To compute visibility of  occluding contours, we adapt the Quantitative Invisibility (QI) algorithm \cite{Appel:1967,BenardHertzmann}.  QI is normally applied to mesh contours; we show how to apply it to rational occluding contours, which requires computing curve intersections and solving equations for cusps efficiently. 


\subsection{Method summary} 

For a given mesh $M$, our method begins with preprocessing steps.  First, we scale $M$ to fit within a unit box. We then compute a global $(u,v)$ parameterization for $M$. We split each triangle into 12 triangles, each of which will be a quadratic patch in the final surface.  Finally, we precompute a factorized matrix that will allow us to efficiently produce patch coefficients for each new viewpoint, minimizing thin-plate surface energy. 

At run-time, given a new camera view $\bc$, we perform the following steps: (1) we apply a projective transformation to the mesh $M$, producing transformed mesh vertices;  (2) we use the precomputed matrices to produce B\'ezier coefficients defining our piecewise quadratic Powell-Sabin  surface from the updated vertex position; (3) we then find all contours contained within all patches; (4) we compute their intersections and cusp points, split them at these points and (5) compute visibility using QI.  The output rational image space curves may be then rendered in 2D. 

Our algorithm allows to perform these steps at a relatively low cost: specifically, 
(2) requires a single backsubstitution solve for a factorized matrix (3) requires solving a quadratic equation and small linear transformations for patches that may contain contours 
(4) requires solving a system of two quadratic equations for cusps for each contour segment, and intersecting  fourth-degree rational curves for a small number of curve pairs, and (5) tracing a small number of rays intersecting them with quadratic patches (also a system 
of two quadratic equations).

%

%% file: figure_tex/patch_contours.tex
\begin{figure}
\centering 
\includegraphics[width=0.4\columnwidth]{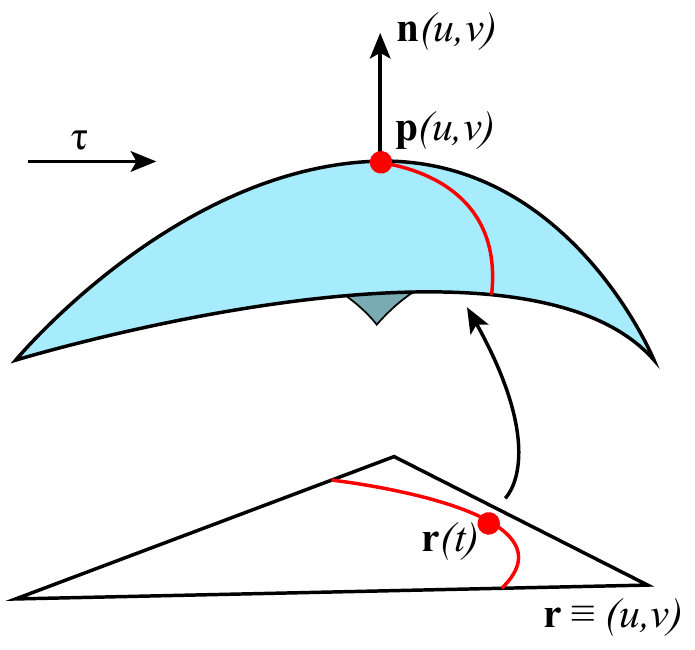}
\includegraphics[width=0.59\columnwidth]{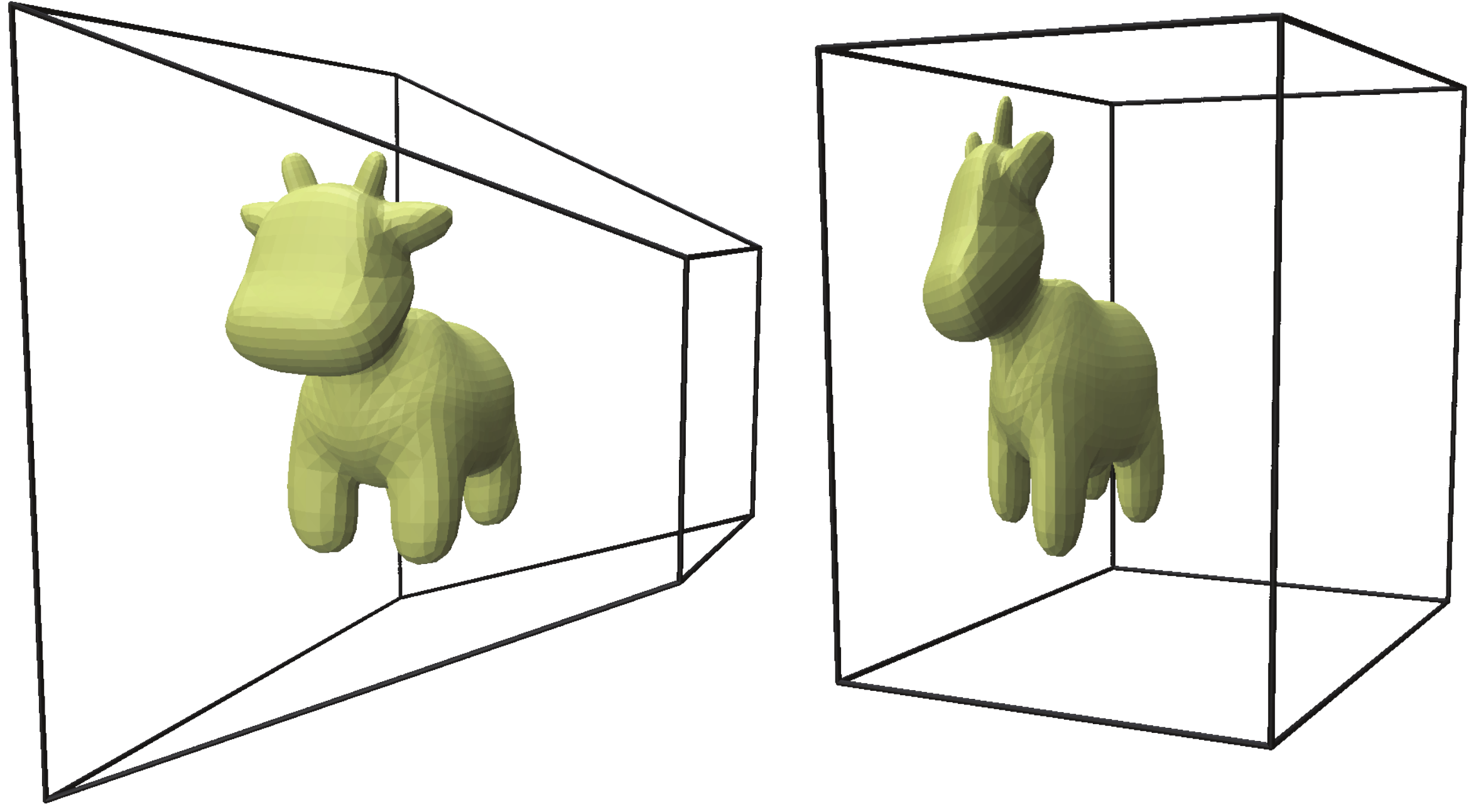}

\caption{Left: The occluding contours of a smooth patch are the points $\bp(u,v)$ where the normal and view direction are orthogonal ($\bn(u,v) \cdot \tau=0$). In this paper, we show that the contour for a quadratic patch can be described by an algebraic curve $\br(t)$ in the parameter domain, so the 3D contour is $\bp(\br(t))$.
Right: 3D perspective transformation converting perspective projection frustum to a cube. (Public domain Spot model by Keenan Crane.)
}
\label{fig:projective}
\label{fig:patch_contours}
\end{figure}

%% file: 35-extraction.tex
\section{Extracting visible contours}

Given a $C^1$ (excluding isolated points) piecewise quadratic surface viewed under orthographic projection, we now  show exact procedures for extracting the occluding contour generator and determining which portions are visible. 
We do assume a general-position view direction, assuming it is perturbed to avoid exact alignment with tangents to flat parts of the surface, and that the surface does not self-intersect.  We do \emph{not} assume that the surface itself is in a fully general position: we \emph{do} require that the surface normal does not vanish, except at cones. 

\subsection{The contours of a single quadratic patch}
\label{sec:quadratic-patch}

We first enumerate parametric expressions for the contours of a quadratic surface, ignoring the triangular domain bounds.

For a quadratic surface parameterized by coordinates $\br=[u,v]$,  the normal $\bn(\br)$ is also quadratic. Hence, the contour equation (\ref{eq:contour-eq}) can be written as 
\begin{equation}
  \tau \cdot \bn(\br) = \frac{1}{2} \br^T A \br +  \bb^T \br + c = 0,
  \label{eq:contour-quadratic}
\end{equation}
i.e.,  the contours are conic sections in the parametric domain (Figure \ref{fig:patch_contours}). For completeness, we now enumerate all stable cases, i.e., the ones that are not always eliminated by a perturbation of the view direction. 

We diagonalize the contour equation using $A=U^T \mathrm{diag}(\sigma_1,\sigma_2) U$.  If $A$ is not singular, then we can reparameterize in terms of unknown $\bz=U(\br-A^{-1} \bb)=[z_1,z_2]$. The contour equation is then:
\begin{equation}
\sigma_1 z_1^2 + \sigma_2 z_2^2 + \hat{c} = 0.
\end{equation}
where $\hat{c}=c-\bb^T{A}^{-1}\bb/2$. We ensure $\hat{c}\geq0$ by negating $\sigma_{1},\sigma_2,\hat{c}$ if necessary. 
We then seek a solution curve of the form $\bz(t)$.
When $\hat{c}>0$, the solution curve $\bz(t)$ may be an ellipse or hyperbola
with scales $k_1 = \sqrt{\hat{c}/|\sigma_1|}$, 
      $k_2 = \sqrt{\hat{c}/|\sigma_2|}$.
In cone patches (Section \ref{sec:surface}), $\hat{c}=0$ occurs stably, in which case the solution $\bz(t)$ is a pair of intersecting lines.
The first three columns of Table \ref{table:cases} provide the solution curves $\bz(t)$ for these cases. Then, the solution curves are converted to parameter domain by $\br(t)=U^T\bz(t) + A^{-1}\bb$.

\input{figure_tex/quadratic_cases.tex}

If $A$ is singular ($\sigma_2=0$), we reparameterize with $\bz=U\br$, and the contour equation becomes
\begin{equation}
\sigma_1 z_1^2 + \hat{\bb}^T \bz + c = 0
\end{equation}
where $\hat{\bb}=U\bb$. 
We negate terms if $\sigma_1<0$.
Then, the solution curve
$\bz(t)$ may be either a parabola or two parallel lines in parameter space, depending on whether $\hat{b}_2=0$. Solution curves $\bz(t)$ are  provided in the fourth and fifth columns of Table \ref{table:cases}.  The parameter-domain curve is then $\br(t)=U^T \bz(t)$.  
We note that $A$ being near-singular can be stable with respect to the viewpoint change for a cylindrical surface;  the contours along a cylinder are always straight lines; see supplemental material for details). 

All tests of a quantity $x$ equal to zero are implemented by checking $|x| < \epsilon$, where we use $\epsilon = 10^{-10}$.

\paragraph{Trimming.}
In general, only a subset of the occluding contour for a quadratic may lie within a patch. We compute the bounds $t \in [t_0,t_1]$ as follows.  The above procedure yields a small number of parametric curves of the form $\br(t)$ that we intersect with each of the lines bounding the domain triangle ($u=0,v=0,u+v=1$). If there are intersections, then the curve is broken into any subintervals $[t_{\min}, t_{\max}]$ contained within the triangle, up to three per curve.


\paragraph{Image-space curve.}
The above steps identify the occluding contour generators within a patch.  Each of these  are rational curves in 3D, given by 
$\bp(\br(t))$, within  bounds $t \in [t_0,t_1]$.
Under orthographic projection with the appropriate rotation, projection to image-space curves amounts to removing the third coordinate from a curve,  yielding 
the final result of the computation, a quartic rational curve.

\subsection{Visibility of contours on a p.w. quadratic $C^1$ surface.}

To compute curve visibility, we adapt the Quantitative Invisibility (QI) algorithm \cite{Appel:1967}.  The QI of a surface point is the number of occluders of that point; a point is visible if and only if it has a QI of zero.   Visibility along a curve can only change at cusps, image-space intersections, and contour-boundary intersections.  Hence, if we split a curve into segments at each of these cases, then QI for an entire segment can be computed by a single ray-test for the curve. Moreover, the number of ray-tests is minimized by propagating QI through these cases, e.g., using the fact that the QI increases by one from the near side to the far side of a cusp. See \cite{BenardHertzmann} for a full description of the QI algorithm.

Previous work applied QI to triangle mesh contours. In order to compute QI for piecewise-quadratic surfaces, we need new algorithms for computing ray tests, detecting cusps, detecting image-space intersections, and propagating QI.

\paragraph{View graph construction.}
We first connect contour generator curves across patch boundaries. Since the surface is $C^1$ by design, the occluding contours will be continuous at patch boundaries, 
and form disjoint loops, with the exception of cone points, where multiple loops can share a point.  We disallow visibility propagation through cone points. 

After we compute cusps and image-space contour intersections as explained below,
we split the countour curves at these points, constructing a graph of non-intersecting  rational quadratic segments with nodes labeled as cusp, intersection, contour-boundary intersection or interior.

\paragraph{Detecting interior cusps.}
To determine cusp positions in the \emph{interior} of a quadratic patch,  we need to find points where $\tau$ is parallel to the tangent to the contour curve. One can use this definition directly, solving the univariate equation $\tau \cdot \frac{d}{dt} \bp(\br(t)) =0 $; however, this equation is a degree 10 polynomial. Instead we opt for a different approach. We introduce two orthogonal vectors 
    $\tau_1$ and $\tau_2$ perpendicular to the view direction, and define the cusp as the point on the contour for which  the tangent is orthogonal to both of these vectors. An important observation is that  the unnormalized tangent to all isolines of  $\bn \cdot \tau$ in the parametric domain  is given by  $[-\bn_v \cdot \tau, \bn_u \cdot \tau]$, which is a linear function on the whole patch.  Mapped to 3D this yields a quadratic equation: 
\begin{align}
    \bt(u,v) =  -\bp_u  (\bn_v \cdot \tau) + \bp_v (\bn_u \cdot \tau)
    \label{eq:tangent}
\end{align}
    i.e., the components of the tangent are also quadratic,
    and the condition for a cusp is a system of two quadratic equations in $(u,v)$:
\begin{align}
    \bt(u,v) \cdot \tau_1 = 0,\; \mbox{and}\; \bt(u,v) \cdot \tau_2 = 0,
\end{align}
Note that the contour equation itself is redundant, because if the view direction is aligned with a surface tangent, it is perpendicular to the normal.
We solve this equation using the pencil method which reduces it to a cubic equation in one variable, and a pair of quadratic equations, which we also use for ray-patch intersections \cite{ogaki2011}; the cubic equation is solved by finding the companion matrix eigenvalues with QR decomposition. 

QI increases by 1 along the contour in the direction of $\bt$.

\input{figure_tex/cusps.tex}


\paragraph{Patch-boundary cusps.}
At a patch boundary, a contour may have two distinct tangents $\bt$ and $\bt'$ corresponding to the common endpoint of the contour segments in two patches; a cusp occurs if $\tau \cdot (\bn \times \bt)$ and $\tau \cdot (\bn \times \bt')$ have different signs. While in this case there is no need to split either of the curves at the cusp,  the visibility may change at the common endpoint, which we refer to as a \emph{boundary cusp}. We opt not to propagate visibility through such cusps. 

\paragraph{Image-space contour intersections.}
The intersections of contours in the image domain are computed using the standard B\'ezier clipping algorithm \cite{sederberg1990curve}, which typically converges in a few iterations. The QI on the top contour is a constant $q$, the segment of the lower contour not occluded by the near surface also has QI $q$, and the lower contour segment occluded by the surface has QI $q + 2$. 

\paragraph{Ray tests and QI propagation.}
Once the contours have been split at cusps and intersections, the visibility can be determined by performing a ray test at a contour segment and then propagating QI. We do not propagate QI through cone points, since multiple contours may meet there, and we do not propagate QI though patch-boundary cusps. If any segments do not have QI values assigned after propagation, the process repeats with a new ray test. We use a highly efficient ray-quadratic patch intersection test \cite{ogaki2011}; however, a substantial speed up is obtained using the standard bounding-box tests and spatial grid for acceleration. 


%% file: figure_tex/quadratic_cases.tex
\begin{table*}[t]
\centering
\caption{
The occluding contours of a quadratic patch under orthographic projection can be parameterized in one of these five forms, depending on the parameters in Equation \ref{eq:contour-quadratic}. Other cases not listed (e.g., $\sigma_1,\sigma_2$>0) do not produce stable contours. See Section \ref{sec:quadratic-patch} for details.
}
    \begin{tabular}{|c|c|c|c|c|}
      \hline
      \multicolumn{3}{|c|}{$A$ nonsingular ($\sigma_1,\sigma_2 \ne0)$} &
      \multicolumn{2}{c|}{$A$ singular ($\sigma_1\ne0,\sigma_2 =0$)} \\ \hline

$\hat{c} > 0$, $\sigma_1,\sigma_2 < 0$ &
$\hat{c} > 0$, $\sigma_1 \sigma_2 < 0$ &
$\hat{c} = 0$, $\sigma_1\sigma_2 < 0$  &
$\hat{\bb}_2 \neq 0$ &
$\hat{\bb}_2 = 0, \hat{\bb}_1^2 >4 \sigma_1^2 c$ \\\hline      
Ellipse: & Hyperbola: & Intersecting lines: 
& Parabola: & Parallel lines: \\
$    \bz(t)=
    \left[ \frac{k_1 (1-t^2)}{1+t^2},
    \frac{2k_2 t}{1+t^2}\right]$
 & 
$    \bz(t)=
    \left[\frac{k_1 (1+t^2)}{t^2-1},
    \frac{2 k_2 t}{t^2-1}\right]$
    & $\bz(t) = \left[ t, \pm t \sqrt{-\frac{\sigma_1}{\sigma_2}} \right]$
    & $\bz(t) = \left[t,
\frac{-\sigma_1 t^2 - \hat{b}_1 t -c}{\hat{b}_2}
\right]
$
    & $\bz(t) = \left[ \frac{-\hat{\bb}_1 \pm \sqrt{\hat{\bb}_1^2 - 4 \sigma_1^2 c}}{2\sigma_1^2}, t\right]$
\\
      \hline
    \end{tabular}
\label{table:cases}
\end{table*}


%% file: figure_tex/cusps.tex
\begin{figure}
\centering
\includegraphics[width=0.8\columnwidth]{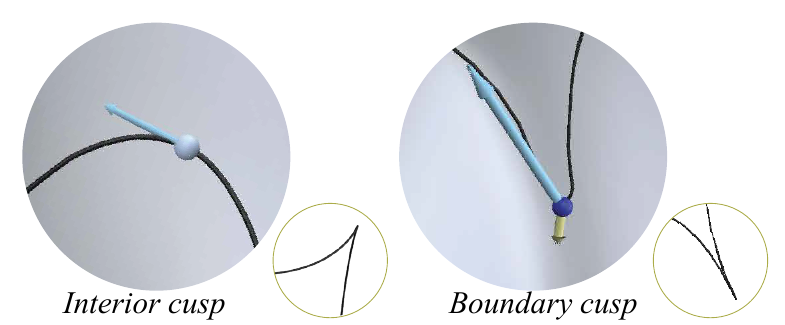}
\caption{ Two types of cusps, a cusp interior to a patch, and a boundary cusp.
}
\end{figure}        

%% file: 40-construction.tex

\section{Surface construction}
\label{sec:surface} 
We now describe our smooth surface construction; producing a high-quality surface is essential to generating well-behaved contours. The input is an arbitrary manifold mesh $M = (V,E,F)$ possibly with boundary, along with vertex positions.  The output is a surface composed of quadratic patches joined with $G^1$ continuity, everywhere except at a small number of isolated vertices.   Our approach is based on He et al.~\shortcite{he2005c}, with several important changes.  




\subsection{Conformal parameterization}
\label{sec:parameterization}

We first compute a locally bijective global surface parameterization, to produce $uv$ coordinates for the mesh, as required for the Powell-Sabin interpolant.
More specifically, the parameterization is a flattening into the $uv$ plane of the input mesh $M$  (Figure \ref{fig:parameterization}). The mesh is cut to a disk $M^c$ along mesh edges, such that,  for two images of a cut edge in the $uv$ domain, their lengths match. The metric is almost everywhere flat if, for all vertices---including all cut vertices, but excluding a small number of \emph{cone vertices}---the sum of the angles in the parametric domain is  $2\pi$. 

\input{figure_tex/parameterization.tex}

The first step is to select a small number of cone vertices, consistent with genus; we use $4(g-1)$ cones for $g > 1$ and  $8$ for genus zero. We uniformly distribute them over the surface, by partitioning it into clusters of close size, and, for each cluster, picking the vertex of smallest discrete curvature, as this minimizes visual artifacts of the surface.  We do not fix angles at the cones; instead, we fix the conformal scale factors at cones to 1, as in \cite{springborn2008conformal}, to minimize parametrization distortion at these points.  Additionally, we perform one step of Loop subdivision on the incident triangles, to further reduce artifacts. 

Then, we apply an efficient implementation of discrete conformal maps \cite{gillespie2021discrete,campen2021efficient} mathematically guaranteed \cite{gu2018discrete} to satisfy these requirements, 
possibly with a moderate amount of mesh refinement.   These methods compute an edge length assignment, which we convert to a $uv$ parametrization using a greedy layout algorithm, as in \cite{springborn2008conformal}.

While conformal parameterization can have extreme scale variation, uniformly scaling the parametric domain  of a polynomial patch does not affect its shape, and so this scaling does not significantly impact our output surface construction.
This stage of our algorithm differs from the parameterization step of He et al.~\shortcite{he2005c}, which does not guarantee injectivity, fixes cone angles to multiples of $\pi$, and does not provide control over their position, leading to higher distortion.

\subsection{Powell-Sabin construction}
The Powell-Sabin interpolant  transforms a set of input degrees-of-freedom (DOFs) $q$
associated with a mesh $M$ into a set of quadratic patches $\bp(u,v)$. Each patch is defined by six B\'ezier points on a triangle in the $uv$ plane.  
Each of these triangles is obtained by splitting the parametric images of the triangles in the original mesh (Figure~\ref{fig:twelve-split}(right)). We use the more-expensive 12-split Powell-Sabin construction as it yields higher surface quality, but a 6-split can also be used with acceptable results.  All quadratic patches associated with an input triangle form a $C^1$ piecewise-quadratic macropatch $\bP(u,v)$. 


The input  coefficients include, for every vertex, the position $\bp$ and two tangents $\bg^u,\bg^v$,  and  for every edge midpoint, a single tangent $\bg^m$ (Figure~\ref{fig:twelve-split}(left)). If the parametrization has no cuts, 
then  $\bg^u$, $\bg^v$ are  prescribed values of ${\partial_u}\bP(u_i,v_i)$ and  $\partial_v \bP(u_i,v_i)$  at a vertex $i$ respectively, and $\bg^m$ is  $\partial_{\be^\perp_{ij}} \bP(u,v)$, where 
$ \be^\perp_{ij}$ is the vector perpendicular to the edge $\be_{ij}$. All macropatches $\bP$ sharing a vertex share these coefficients at the vertex, and similarly, when two patches sharing a midpoint, they share a coordinate derivative in the direction $\be^\perp_{ij}$. Powell and Sabin~\shortcite{powell1977piecewise} show that sharing these degrees of freedom is sufficient for their construction to yield a $C^1$ surface.  

We write the complete set of free DOFs for the surface as a vector of coefficients:
\begin{align}
q &= [\bp_1, \bg^u_1, \bg^v_1  \ldots \bp_n, \bg^u_n, \bg^v_n, \bg^m_1 \ldots \bg^m_m]  
\end{align}
where $n$ is the number of vertices, $m$ the number of edges, and $\bg^m_\ell = \bg^m_{ij}$ if the edge $(ij)$ has index $\ell$; the 3 components of each vector are 
flattened. 

\paragraph{Local parameters.}
To construct a single macropatch for a triangle $(ijk)$ we extract the relevant 12 DOFs from  this vector to obtain a \emph{local} DOF vector, $q^{\mathrm{loc}} = 
[\bp_i, \bg^u_i, \ldots \bg^m_k]$, with 12 vector DOFs; the coefficients 
of the 12 quadratic subpatches are obtained by  the standard Powell-Sabin linear transformation of this vector, described in supplementary: this yields six Bezier points for each quadratic patch as shown in Figure~\ref{fig:twelve-split}.
\[
 p^{\ell} = B_{\ell} q^{\mathrm{loc}},\; \ell = 1 \ldots 12
\]
where each $B_\ell$ is a $6 \times 12$  matrix applied coordinatewise, mapping twelve local degrees of freedom to six quadratic patch coefficients. 

\paragraph{Cut vertices.}
To determine local DOFs  $q^{\mathrm{loc}}$ from global $q$ at a vertex $i$ or edge $(ij)$ on the cut, we need to define the local gradients $\bg$ consistently across the cut. We do so by constructing a chart comprising all incident triangles of the vertex or edge.   One of the edges of the chart is arbitrarily chosen to be the $u'$ coordinate direction, defining a local coordinate system $(u',v')$. 
All triangles are mapped to the chart by a rigid transformation $T_{ijk}$ from global $\br = (u,v)$  to chart coordinates $\br' = (u',v')$ (Figure~\ref{fig:parameterization}(c)).
Constructing such a common domain is possible because the parameterization (Section \ref{sec:parameterization}) produces angles that sum up to $2\pi$ and 
edge lengths that match across the cut.  

Then we can obtain the gradient DOFs $\bg$ in coordinates  $(u',v')$ on the chart. 
To obtain the vertex DOFs for each triangle $(ijk)$,   we apply $T_{ijk}$ and apply the corresponding transform to $\nabla_{\br'} \bP(\br') =[\partial_{u'} \bP, \partial_{v'} \bP]$, to obtain the local DOF vector $\nabla_{\br} \bP = T_{ijk} \nabla_{\br'} \bP$ for the macropatch.  This construction retains $C^1$ continuity  across cut edges and at non-cone vertices \cite{he2005c}. 
  


\input{figure_tex/twelve-split.tex}



\begin{figure}
\includegraphics[width=0.8\columnwidth]{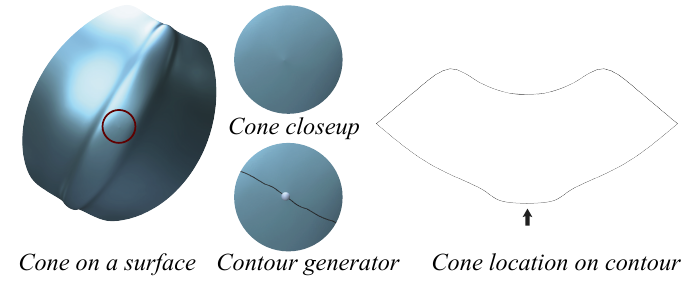}
\caption{Surface and contour behavior near a cone. (Open source Helmet model from OpenSubdiv.)}
    \label{fig:cone}
\end{figure}

\paragraph{Cone vertices.}
At cone vertices, we create degenerate patches, e.g., \cite{neamtu1994degenerate}.  In order to achieve consistent gradients across the cut, we set  gradients $\bg_u, \bg_v$ to zero at cone vertices, making the patch map $\bp(u,v)$ singular at these points.  The conditions for $C^1$ continuity are still satisfied, although the surface has a cone point at the vertex (Supplemental Material). 
While one could apply spline surface construction methods that force a common tangent plane at a  singular vertex, we found our approach to have negligible impact on contour quality. 
Typical surface behavior near a cone is shown in Figure~\ref{fig:cone}.

\subsection{Thin-plate optimization}  
We now describe how to optimize the global parameters $q$ to produce a smooth surface shape that approximates the input mesh vertices.
One nice property of a Powell-Sabin spline surface 
is that it is $C^1$, and, at the same time, minimal degree.  $C^1$ is a requirement for \emph {conforming} finite-element discretization of optimization objectives containing second derivatives, in particular, the thin-plate functional:
\begin{align}
  E(q)&=  \sum_\ell \int_{\Omega_\ell} \bp_{uu}^2 + \bp_{vv}^2 + 2\bp_{uv}^2 du dv + w \sum_i A_i (\bp_i - \bp_i^0)^2
    \label{eq:integral-objective}
\end{align}
where $\bp_i^0$ are the input mesh vertices and $\Omega_\ell$ is the $\ell$-th triangle domain. 
We discretize the functional by simply substituting the expressions for quadratic patches
in terms of the degrees of freedom  $q^{loc}$, yielding constant expressions for the integrands for each quadratic patch. These expressions are quadratic in the degrees of freedom, so the energy can be written as a quadratic form for each patch with 
a constant local matrix,  which are then assembled into a global system.  We pre-compute  matrices $H^s$ and $H^f$, combined into a matrix  $H = H^s + w H^f$, corresponding to the fitting and smoothing terms of the quadratic objective  \eqref{eq:integral-objective} respectively.  Then the energy can be written in the form 
\begin{align}
E(q) &= \frac{1}{2} (q^T H^s q  +  w (q-q_0)^T H^f (q-q_0)) \\ &= 
\frac{1}{2} q^T H q  -  w q^T H^f q_0 + \mathrm{const} 
    \label{eq:objective}
\end{align}
The constant term $q_0$ is the vector of degrees of freedom with 
the original mesh vertex positions for $\bp_i$ and zeros for all other components.
This objective is minimized by solving $Hq   = w H^f q_0$ for $q$, i.e. a single sparse linear solve.   

One can also show by direct computation that our optimization objective minimum, expressed as B\'ezier control points of patches, is independent of the choice of the chart coordinates $(u',v')$.

\paragraph{Efficient computation.}
Changes to viewpoint change the mesh vertex positions due to the projective transformation (Section \ref{sec:overview}), thus changing $q_0$.  However, the matrix $H$ depends on parametric coordinates and \emph{not} on the vertex positions. Hence, all of the computations in this section, including mesh parameterization and defining local and global coordinates, can be computed in a preprocess. We precompute the Cholesky decomposition of the sparse matrix $H$,
which makes the cost of solving the system  $Hq =  w H^f q_0$ negligible at run-time.





%% file: figure_tex/parameterization.tex
\begin{figure}
    \centering
    \includegraphics[width=0.8\columnwidth]{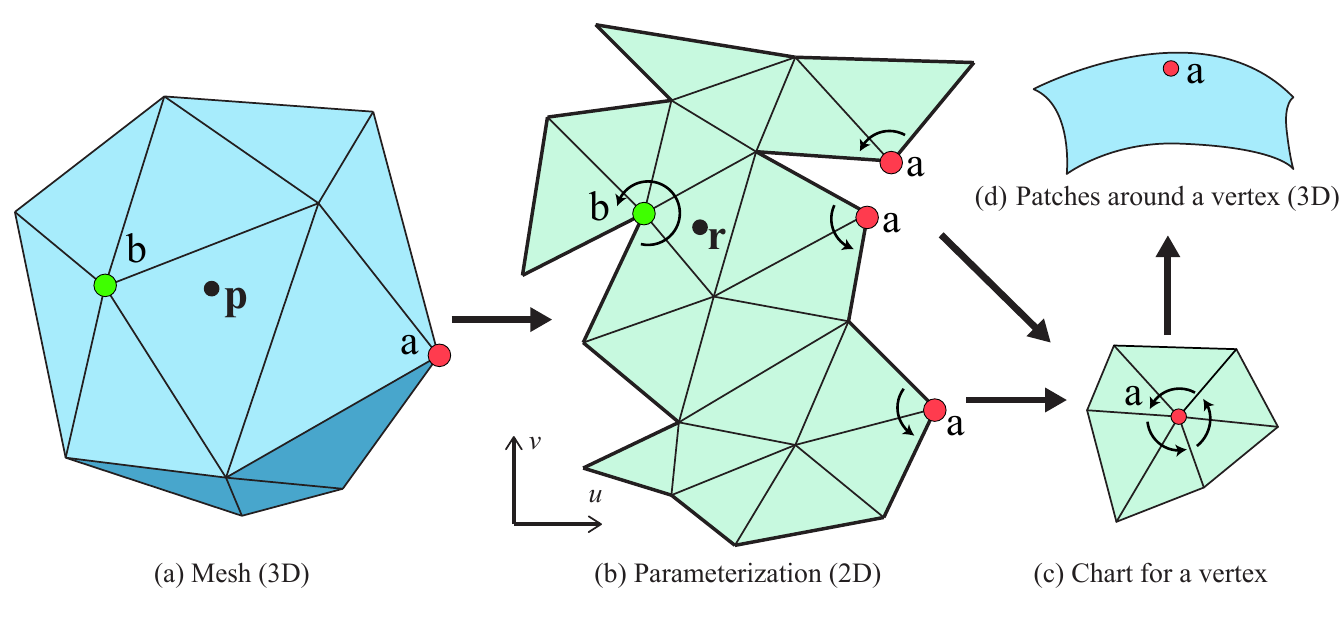}

    \caption{Surface definition.
    (a) An input mesh with vertex positions $\bp_i$.
    (b) 3D mesh is mapped to the plane with cuts with texture coordinates $\br=(u,v)$, consistent edge length and sum of angles at most vertices equal to $2\pi$  (c) Vertex $a$ (red vertex) occurs at three places on the cut, but a chart can be constructed where the sum of the angles around $a$ is $2\pi$. For a cone vertex $b$ (green vertex), the sum is less than $2\pi$.
    (d) The chart for each vertex with respect to which gradient DOF are defined. 
    }
    \label{fig:parameterization}
\end{figure}

%% file: figure_tex/twelve-split.tex
\begin{figure}
    \centering
\includegraphics[width=0.3\columnwidth]{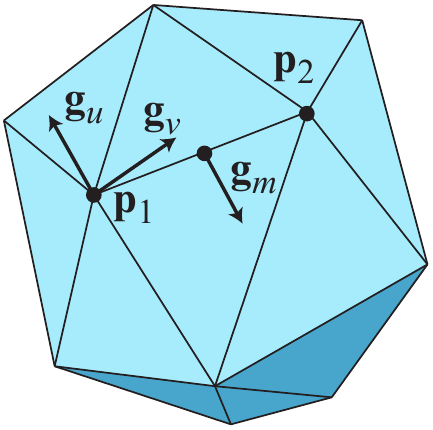}
    \includegraphics[width=0.4\columnwidth]{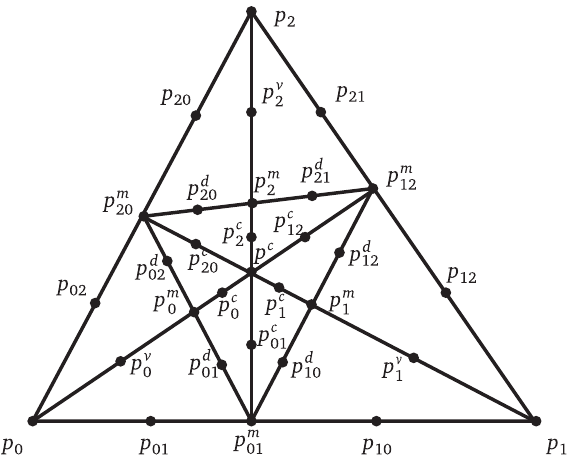}
    \caption{
    Left:    The smooth surface is parameterized by the following coefficients: for each vertex, a position $\bp$ and gradients $\bg_u$ and $\bg_v$, and, for each edge, a midpoint gradient $\bg_m$.
    Right: Each input triangle is split into 12 triangles, with B\'ezier controls named as shown here. These control points are computed from the smooth surface coefficients $q^{\mathrm{loc}}=[\bp_i, \bg_i^u, ..., \bg^m_k]$.
    }
    \label{fig:twelve-split}
\end{figure}

    

%% file: 50-results.tex
\section{Evaluation}
\label{sec:results}

We show results of our method in Figures \ref{fig:teaser} and \ref{fig:results}. Our method produces clean, smooth rational output curves without gaps or other topological errors.

\paragraph{Timings and scaling.} We tested our algorithm on the same small test set that was used in \cite{ConTesse}.  We preprocessed the meshes by performing boolean unions of parts to eliminate all intersections, and cleaning up the resulting mesh to eliminate very short edges.  Figure~\ref{fig:timing} shows the timings for  the view-dependent and  view-independent parts of our algorithm. We use 29  models and 26 randomized views per model. We see relatively very little per-frame timing variation (standard deviation less than 150~ms for the largest models), and both precomputation and per-view performance scales approximately linearly with the input size. Our implementation is serial and not heavily optimized, and we expect that performance can be significantly improved.


\paragraph{Comparison with previous work.}  
Precise comparison with ConTesse \cite{ConTesse}, the state-of-the-art method for accurate contours, is difficult for various reasons, e.g., that paper reports timing only for mesh generation. Nonetheless, it is clear that our approach operates an order of magnitude faster during run-time, and scales much better as well.  Our entire per-view processing time, for smaller meshes, is comparable to ConTesse's mesh generation step alone. However, for larger meshes, our method becomes orders-of-magnitude faster, e.g., for ``Fertility,'' ConTesse averages 16 seconds per view for mesh generation, whereas we require 0.4 seconds per view for the entire visibility pipeline, after a 4.4 seconds of preprocessing time. For "Killeroo," ConTesse requires 33 seconds per view for mesh generation, whereas we require 0.5 seconds per view, after 3 seconds preprocessing.
We evaluated timing on an MacBook Pro 2.7 GHz, an older machine than the ConTesse results are reported on.

This is to be expected, since our method requires only a linear solve to compute patch coefficients for each new view, whereas ConTesse employs many iterative heuristic search steps to find a valid mesh for each new view.  


\paragraph{View-dependent effects.}  Because we construct a view-dependent surface approximation, in principle, the object could appear non-rigid during rotation.
Figure~\ref{fig:projective-fov} and the accompanying video show that the projective transformation does not alter the geometric appearance or produce non-rigid effects.

\paragraph{Smoothness vs.~approximation.} All examples in this paper were generated using a constant weight $w=1$ for the fitting term in \eqref{eq:objective}; Figure~\ref{fig:fitting-weight} shows how the surface changes as this weight varies.



%% file: 60-concl.tex
\section{Conclusions}
\label{sec:concl}

We have presented a method for efficient computation of high-quality occluding contours on $C^1$ surfaces approximating arbitrary input meshes. 
As contours are computed in closed algebraic form and are the exact (modulo numerical errors) contours of the surface, visibility computation is a straightforward extension of the QI algorithm for meshes, and, at the same time, are close to the contours on the smooth surface that the mesh is approximating. The resulting contour lines cannot violate topological conditions from \cite{ConTesse}, except due to numerical error for non-general-position choices of the view direction.  Hence, the output contours have valid, well-defined visibility, without the artifacts that have plagued previous methods.

There are many extensions that are easy to add to our framework:  it is straightforward to add sharp features, and handle non-manifold surfaces, e.g., resulting from intersections or self-intersections. Other curves, like  suggestive contours, apparent ridges \cite{DeCarlo:2012}, and can be added as polylines, as well as stylized shading effects. The per-view computational cost of our approach is proportional to the number of contour segments  and is embarrassingly parallel: an optimized implementation is likely to improve performance by a large factor.  
Another potentially important direction is to explore higher-order constructions with constraints with reducible parametric contour equations, as these may eliminate the need for a global parametrization.



%% file: 65-figures.tex
\newpage

\input{figure_tex/results-figure.tex}

\begin{figure*}
    \centering
    \includegraphics[width=5in]{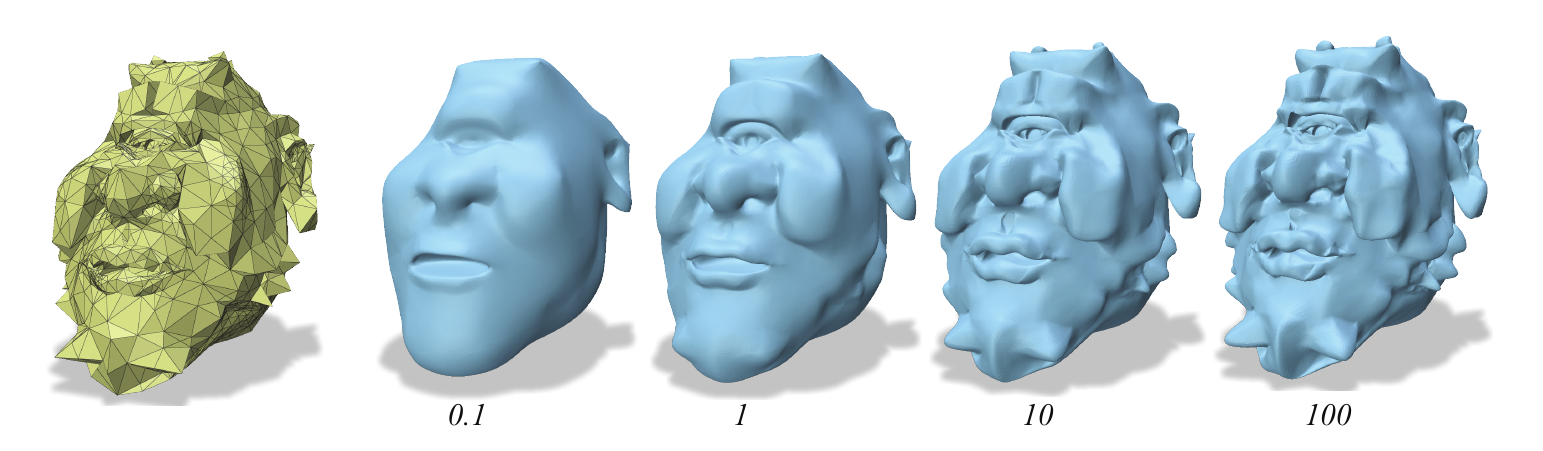}
    \caption{Given a triangle mesh, we need a representation of the underlying smooth surface.  Here we visualize the dependence of our smooth surface shape fitting on the weight $w$, without a projective transformation step. We use $w=1$ for all experiments in this paper. (Public domain Ogre model by Keenan Crane.)}
    \label{fig:fitting-weight}
\end{figure*}

\begin{figure*}
    \centering
    \includegraphics[width=5in]{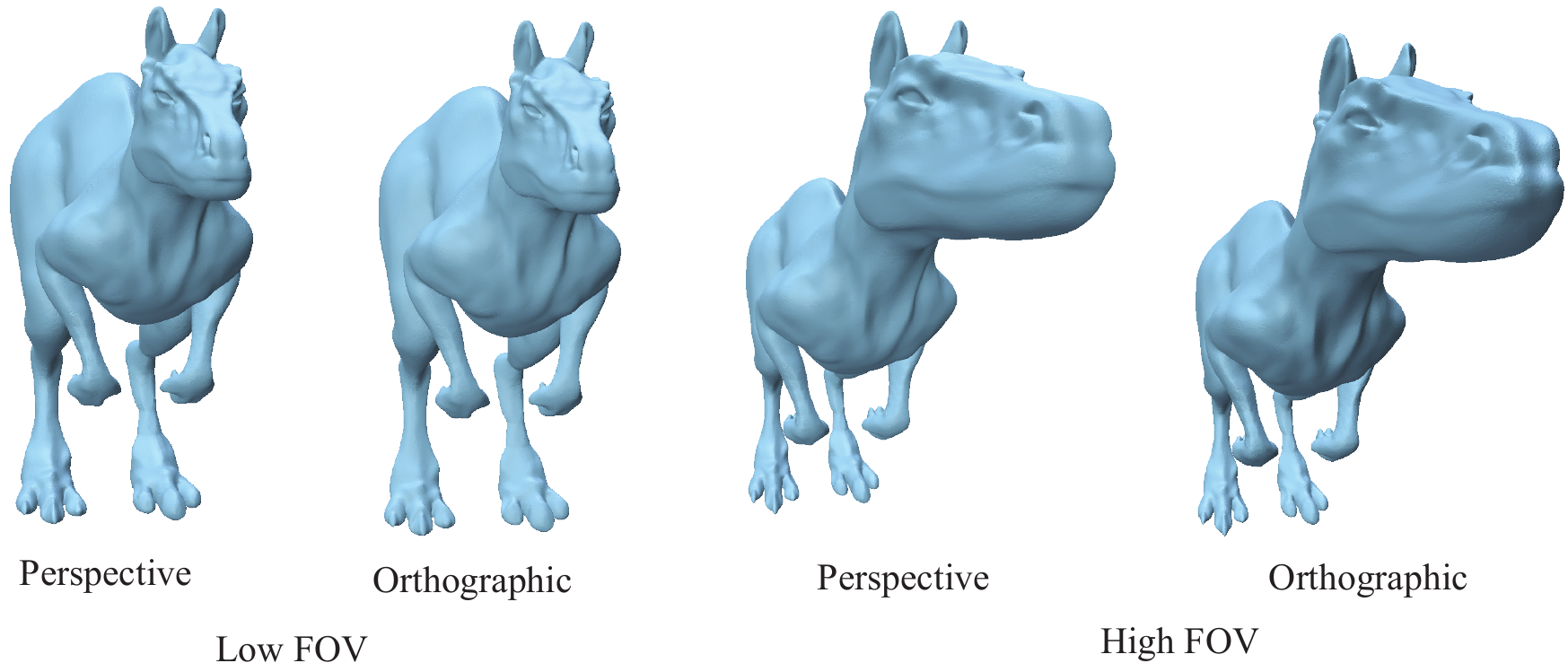}
    \caption{Perspective projection of a mesh, vs.~orthographic projection of the projective-transformed mesh. In each pair, the two renderings are geometrically equivalent. (The shading differs, but is not used in contour detection and shown here only for visualization. Killeroo \copyright headus.com.au.) }
    \label{fig:projective-fov}
\end{figure*}

\begin{figure*}
\centering
\includegraphics[width=5in]{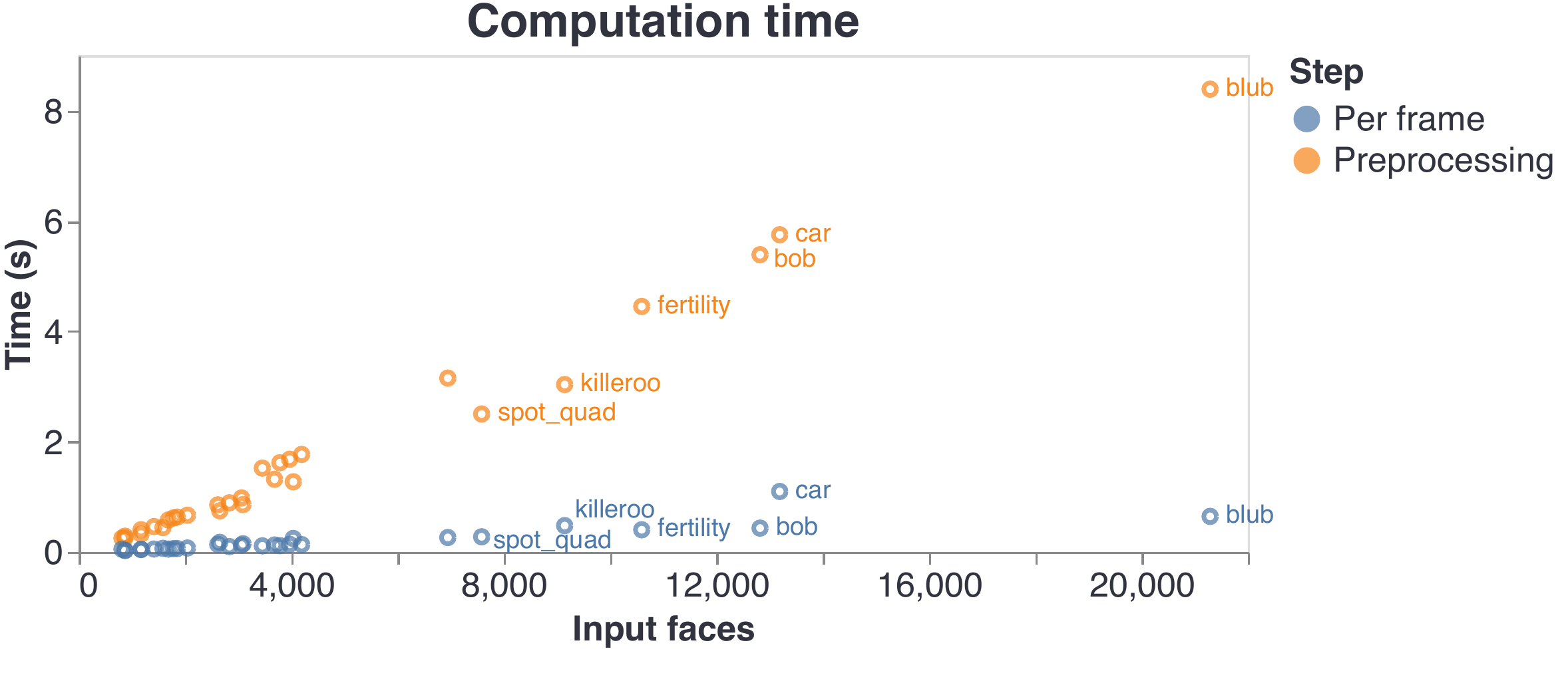}
\caption{Timing for our method, on meshes from the ConTesse \cite{ConTesse} dataset. Orange circles show preprocessing computation time, and blue circles show per-frame time, averaged over 26 frames. While preprocessing can take several seconds, per-frame computation is very efficient, growing very slowly as a function of mesh size. Computation times were measured on a MacBook Pro 2.7Ghz Intel Core i7, 16Gb memory. The "car" mesh is more intensive at run-time due a very high shape complexity, requiring over a million ray-patch intersection tests per frame.
}
\label{fig:timing}
\end{figure*}

\newpage

%% file: figure_tex/results-figure.tex
\begin{figure*}
    \centering
    \includegraphics[height=0.95\textheight]{images/fig-example-paper.pdf}
    \caption{
       Examples of of contour line sets obtained using our method on models from the
       modified dataset of \protect\cite{ConTesse}. (Bigguy and Monster Frog \copyright Bay Raitt. Fertility courtesy UU from AIM@SHAPE-VISIONAIR Shape Repository. Public domain Blub, Bob, and Ogre models by Keenan Crane. Killeroo \copyright headus.com.au. Open source Pawn model from OpenSubdiv.)
    }
    \label{fig:results}
\end{figure*}

%% file: 70-suppl.tex


\section{Conic degenerate patches}

A quadratic  Bezier patch  can be written as 
\[\bp(u,v) = \bp_{00} w^2 +    \bp_{20} u^2 + \bp_{02} v^2    + 2\bp_{10} uw  +  2\bp_{01} v w + 2\bp_{11} v u,\] with  
the first three control points corresponding to the vertices and the last three to the midpoints of a Bezier triangle, and $w=1-u-v$. 
Suppose the gradient at $\bp_{00}$ is zero; then then the equation  \eqref{eq:globtoloc} and formula  for the Bezier points of subpatches below 
yield $\bp_{01} = \bp_{10}  = \bp_{00}$, i.e., the three points are collapsed to one, and the patch reduces to 

\[
\bp(u,v) = \bp_{00} w (w + 2v + 2u) +  \bp_{20} u^2 + \bp_{02} v^2  +  2\bp_{11} v u
\]
A direct computation yields the following expressions for the tangents: 

\[
\bp_u =     2(\Delta \bp_{20} u + \Delta \bp_{11}v), \; \bp_v =     2(\Delta \bp_{02} v + \Delta \bp_{11} u)
\]
where $\Delta \bp_{ij} = \bp_{ij}-\bp_{00}$. From these, we can see that the tangent direction for any line  $a u + b v = 0$ is constant, i.e., the 3D images of these lines, passing through $\bp_{00}$, are straight lines, and the patch is a cone.  For these patches, the contour lines are stably a pair of intersecting lines.  We also note that if  the surface is tangent plane continuous at such a point, it follows that the three vectors  $\Delta \bp_{ij}$ have to be coplanar in general, i.e., all control points of the patch are in the same plane and the patch is flat. 
We found that forcing such flat spots around these vertices is often less preferable compared to allowing a cone vertex, in terms of contour generator behavior. 

\section{Powell-Sabin interpolant} 

We describe the construction for a scalar function. Exactly the same construction is applied to each of $x$, $y$ and $z$ coordinate functions. 
For completeness, we describe the Powell-Sabin interpolant construction for the global parametrization setting in which the gradient degrees of freedom per vertex are defined on vertex charts. 
As explained in the text,  first, for each corner of a triangle $(i,j,k)$  transforms $T_{ijk}$ are applied to transform the gradients $g^u$, $g^v$ defined in the coordinate charts for each vertex 
$i,j,k$, and $g^m$ defined in local coordinates on an edge chart, to the global coordinates $u,v$ in which the parametric coordinates $\br_i,\br_j,\br_k$ of the triangle are defined.  We proceed with defining the set of quadratic patches as follows.  

Given these 12 degrees of freedom transformed to the global coordinates, we first convert them to an affine-invariant form of derivatives per patch.  

We switch from global triangle vertex indices $(i,j,k)$ to local indices $(0,1,2)$.  The per-triangle local degrees of freedom are: 
(1) 3 vertex values $p_i$, $i=0,1,2$, (2) 6 derivatives of 
$p(u,v)$  at vertices $i$ in the directions of parametric edges $e_{ij}$, which we denote $d_{ij}$, and $h_{ij}$, (3) 3 derivatives of $p(u,v)$ evaluated at midpoints $m_{ij}$ of the edges, in the direction of the opposite vertex $k$, where $(i,j,k)$ is a permutation of $(0,1,2)$.  

The transformed vectors $(g^u_i, g^v_i)$ are derivatives of the quadratic interpolant in directions $u$ and $v$, and $g^m_{ij}$ are derivatives along the direction 
$\be_{ij}^\perp$.  The derivative $g^{e}_{ij}$, along $\be_{ij}$ is fully determined from other degrees of freedom by the requirement that the quadratic patches inside each triangle join with $C^1$ continuity: $g^{e}_{ij} = -2p_i - d_{ij}/2 + 2 p_j + d_{ji}/2$.  This leads to the following  linear map for the triangle local degrees of freedom (note that the coefficients of the equations depend on the parametric coordinates only, so are view-independent). 

\begin{equation} 
d_{ij} =  g^u_i e^u_{ij} + g^v_i e^v_{ij},\quad 
h_{ij} =   g^e_{ij} \be^m_{ij} \cdot \hat{\be}_{ij}  + 
g^{m}_{ij} \be^m_{ij} \cdot \hat{\be}_{ij}^\perp 
\label{eq:globtoloc}
\end{equation}
where $\be^m_{ij} = \br_k - \frac{1}{2}(\br_i + \br_j)$ is the vector from the midpoint to the opposite vertex in the parametric domain. 

\paragraph{Bezier points for subpatches.}
Finally, given the degrees of freedom above, the Bezier control points of each of the 12 quadratic patches denoted as shown in Figure~\ref{fig:twelve-split} in a simple form: the corner control points are just $p_i$,  adjacent six points $p_{ij}$ are displaced proportionately to edge derivatives $d_{ij}$; $h_{ij}$ determines the displacement of $p^c_i$ from the midpoint of $(p_{ij},p_{ij})$ and the rest are determined by averaging.  Note that the coefficients of these expressions are constants not depending on the parametrization,  so this yields a fixed part of the matrix, with the 
dependence on the parametrization entirely captured by \eqref{eq:globtoloc}

\begin{center}
\begin{tabular}{r@{\hspace{1mm}}lr@{\hspace{1mm}}l}
$p^c_i $&$= \frac{1}{2}(p^c_{ij} + p^c_{ki})$ \\
$p_{ij}$&$= p_i + \frac{d_{ij}}{4}$ & $p^v_i $&$= \frac{1}{2}(p_{ij}+p_{ik})$ \\
$p_{ij}^m$&$ = \frac{1}{2}(p_{ij}+p_{ji})$& $p_i^m $&$=  \frac{p^v_i}{4} + \frac{3p^c_i}{4}$\\
$p_{ij}^c$&$ = p^m_{ij} + \frac{h_{ij}}{6}$ & $p_{ij}^d $&$=\frac{p_{ij}}{4} + \frac{3p^c_{ij}}{4}$ \\
$p^c$&$ = \frac{1}{3}(p^c_{01} + p^c_{12} + p^c_{20})$ & 
\end{tabular}

\begin{figure}
    \centering
    \includegraphics[width=\columnwidth]{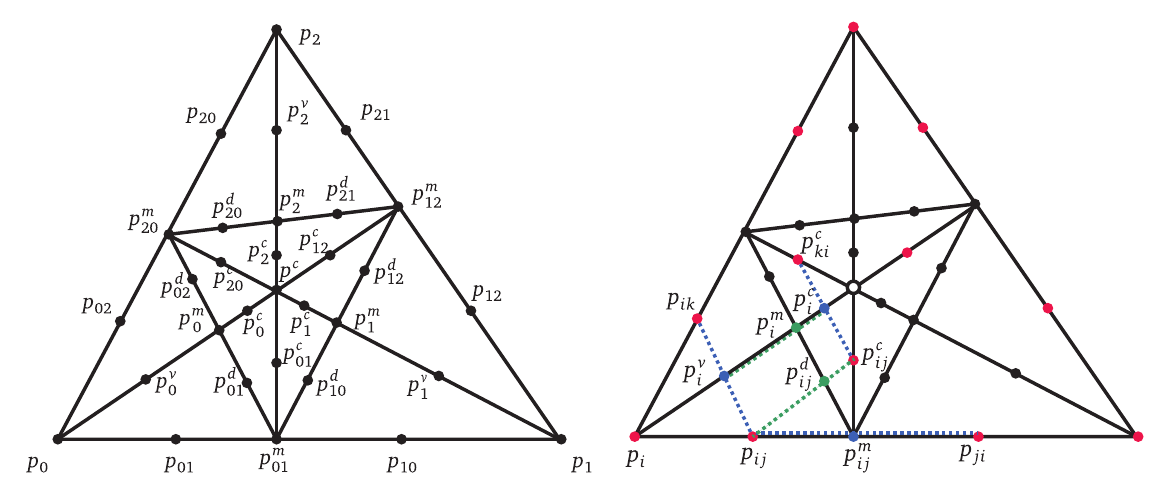}
    \caption{Left: names of control points for the Bezier representation of the patches in the 12-split Powell-Sabin interpolant. Right: red points indicate control points computed directly from the input data; blue points are computed by averaging of pairs of points connected by the blue dotted lines. green points are obtained by averaging points at the ends of green dotted lines with weights $3/4$ and $1/4$. 
    }
    \label{fig:twelve-split}
\end{figure}

\end{center}